\documentstyle[aps,prb,manuscript,eqsecnum]{revtex}

\begin{document}

\title{A simple semiclassical approach to Kramers' problem}

\author{ Jyotipratim Ray Chaudhuri, Bidhan Chandra Bag and 
Deb Shankar Ray}

\address{Indian Association for the Cultivation of Science, Jadavpur,
Calcutta 700 032, INDIA.}

\maketitle

\draft

\begin{abstract}
We show that Wigner-Leggett-Caldeira equation for Wigner phase space 
distribution function which describes the quantum Brownian motion of a 
particle in a force field in a high temerature, Ohmic environment can be
identified as a semiclassical version of Kramers' equation. Based on an 
expansion in powers of $\hbar$ we solve this equation to calculate the 
semiclassical correction to Kramers' rate.
\end{abstract}

\vspace{2.0cm}



\section{Introduction}

The recent two decades have witnessed a phenomenal development of the theory 
of noise induced escape from metastable states. The escape is governed by
Brownian motion in addition to the characteristic dynamical motion
of the system in presence of the potential field, $V(x)$, Brownian motion
being driven by thermal forces which, in turn, are associated with the 
dissipation $\gamma$ through the fluctuation-dissipation theorem at finite
temperature $T$. The problem \cite{kramers}
has been addressed by a large number of workers
both from classical and quantum mechanical point of view at various levels of 
description under a variety of reasonable approximations in the weak, strong
and intermediate damping limits. The rate theory thus constitutes a 
well-developed body of literature by now \cite{rmp1,vim}.

Although a number of interesting approaches to quantum theory of dissipative 
rate processes based on dynamical semigroup method for evolution of density
operator were proposed \cite{louisell,gg1}
in seventies to treat the quantum and nonlinear
optical phenomena with considerable success, the method could not gain good 
ground in the theory of rate processes due to the fact that it is based on
weak system-reservoir coupling ($\gamma \ll \omega$). The method which got 
major appreciation afterwards in eighties and nineties 
in the wide community of physicists 
and chemists is the real time functional integrals. This method has shown 
to be most effective in treating quantum transition states  
\cite{miller1,voth,liao}, dissipative coherence
quantum effects as well as the incoherent quantum tunneling processes
\cite{rmp1,vim,rmp2,grabert,leggett}. The latter processes
being more relevant in the present context. Notwithstanding
its phenomenal success, it may, however, be noted that compared to classical
Kramers' theory the method of functional integrals for calculation of rate
rests on a fundamentally
different footing. While the classical theory is based on the differential
equation for evolution of probability distribution function of a particle 
executing Brownian motion in a force field, path integral methods
rely on the evaluation of quantum partition function of the
system interacting with a bath of harmonic oscillators with a 
characteristic distribution of frequencies \cite{leggett}. 
The question is whether there is
any natural extension of the classical method to the quantum domain or in
other words what is the analogue of Kramers' equation in the quantum regime.
For this one has to look for an equation for evolution of a  quantum 
probability distribution function of a particle in a force field driven by 
quantum Brownian motion.
Unfortunately the theory of quantum Brownian motion in terms of phase space
probability function \cite{leggett} is far from complete. Under restricted 
conditions one
may use either weak damping limit ($\gamma \ll \omega$) or take resort to
semiclassical procedure ($k_B T \gg \hbar \omega$) and arbitrary damping.
Thus, strictly speaking, to the best of our knowledge, there exist no full
quantum analogue of Kramers' equation valid for arbitrary damping and
temperature.

Our aim in this paper is two-fold :

i) Although there exists no full quantum analogue of Kramers' equation, we
enquire whether there is any semiclassical Kramers' equation. We show that
the Wigner-Leggett-Caldeira [WLC] equation [ see Eq.(\ref{wlc}) ]  for 
Wigner probability phase
space function which describes the quantum Brownian motion of the particle in
a force field in a high temperature Ohmic environment may be interpreted as a 
semiclassical Kramers' equation. WLC equation reduces to classical Kramers'
equation in the limit $\hbar \rightarrow 0$.

ii) While the existing methods of calculation of quantum Kramers' rate are based
on path integral technique, we solve the semiclassical Kramers' equation
for barrier crossing dynamics as a boundary value problem to calculate the
leading order ($\hbar^2$-order) quantum correction to classical rate. WLC
equation had been used earlier in connection with quantum decoherence 
\cite{zurek} and other problems \cite{bag1,bag2}. To the best of our knowledge 
the implementation of a differential equation-based approach for the calculation 
of quantum Kramers' rate like the present one had not been tried upto date.

The outlay of the paper is as follows. In Sec.~{II} we discuss the quantum 
dynamics of a dissipative system and show that a WLC
equation for Wigner probability density function may serve as a good
description for noise-induced escape rate in the semiclassical regime. In
Sec.~{III} elementary formulas are summarized. We apply the appropriate
boundary conditions and make use of an $\hbar$-expansion to solve the
differential equation for stationary probability function in Sec.~{IV}. In
Sec.~V we calculate the steady state flux over the barrier, population
inside the well and the escape rate and show that the quantum contribution
appears in the order $\hbar^2$. The paper is concluded in Sec.~{VI}.


\section{Quantum dynamics of a dissipative system}

We consider a dynamical system characterized by a potential $V(x)$ coupled to an 
environment. The latter is modeled as a set of harmonic oscillators with 
frequencies $\{ \Omega_j \}$. Evolution of such an open quantum system
has been studied over the last several decades under a variety of reasonable 
assumptions \cite{rmp1,vim,louisell,gg1,leggett}. Specifically interesting is 
the semiclassical limit of a high temperature Ohmic environment \cite{leggett}. 
The dissipative time evolution of the Wigner distribution function $W(x, v, t)$ 
for the system with unit mass ($m = 1$) for arbitrarily strong damping in
this case can be described by \cite{leggett}
\begin{equation}
\label{wdf}
\left(\frac{\partial W}{\partial t}\right)_{dissipative} = 2\gamma 
\frac{\partial (v W)}{\partial v}  
+ D \frac{\partial^2 W}{\partial v^2}  \; \;,
\end{equation}

where $\gamma$ and D are the dissipation constant and the diffusion 
coefficient, respectively. The microscopic structure of $\gamma$ is given in
Ref.\cite{leggett}. $x$ and $v$ are c-number co-ordinate and 
velocity variables. The first term in Eq.(\ref{wdf}) 
is a direct consequence of the 
existence of $\gamma$-dependent term in the imaginary part of the exponent
in the expression for the propagator for the density operator in the
Feynman-Vernon theory and has been shown to be responsible for appearance of a
damping force in the classical equation of motion for the Brownian particle
to ensure quantum-classical correspondence. $\gamma$ and D are related by the
fluctuation-dissipation relation, $D = \gamma \hbar \omega
\coth{\frac{\hbar \omega}{2 k_B T}}$ (in the semiclassical 
limit  $D \rightarrow 2 \gamma k_B T$ and $W(x,v,t)$ reduces to classical
phase space distribution function). $\omega$ is the renormalized linear 
frequency of the nonlinear system.
The quantum dynamics of the bare system on the other hand is governed by the
celebrated Wigner equation \cite{wigner} as given by
\begin{equation}
\label{weq}
\left(\frac{\partial W}{\partial t}\right)_{dynamical} = 
-v \frac{\partial W}{\partial x}
+ \frac{\partial V}{\partial x} \frac{\partial W}{\partial v} 
+ \sum_{n\geq 1} 
\frac{ \hbar^{2n}(-1)^n}{ 2^{2n}(2n+1)! }
\frac{\partial^{2n+ 1}V}{\partial x^{2n+ 1} }
\frac{ \partial^{2n+1} W }{\partial v^{2n+1} }  \; \; .
\end{equation} 

The above $c$-number description (\ref{weq}) is equivalent to full 
Schr\"odinger equation. The quantum correction to classical Liouville motion 
is contained in the $\hbar$-containing terms in the sum. We emphasize here 
that  while (\ref{wdf}) is a {\em semiclassical Brownian dynamics} due to the 
coupling of an Ohmic environment to the system, Eq (\ref{weq}) takes into 
account of the full {\em quantum nature} of the system. The overall dynamics 
is then given by
\begin{eqnarray*}
\left(\frac{\partial W}{\partial t}\right) =  
\left(\frac{\partial W}{\partial t}\right)_{dynamical} 
+\left(\frac{\partial W}{\partial t}\right)_{dissipative} \; \; ,
\end{eqnarray*}

\noindent
or

\begin{equation}
\label{wlc}
\frac{\partial W}{\partial t} =
-v \frac{\partial W}{\partial x}
+ \frac{\partial V}{\partial x} \frac{\partial W}{\partial v} 
+ \sum_{n\geq 1} 
\frac{ \hbar^{2n}(-1)^n}{ 2^{2n}(2n+1)! }
\frac{\partial^{2n+ 1}V}{\partial x^{2n+ 1} }
\frac{ \partial^{2n+1} W }{\partial v^{2n+1} } 
+ 2\gamma \frac{\partial (v W)}{\partial v}  
+ D \frac{\partial^2 W}{\partial v^2} \; \; .
\end{equation}

The WLC equation, (\ref{wlc}) is the starting point of 
our further analysis. Before proceeding further several points are to be 
noted :

First, strictly speaking $\gamma$ and D terms in Eq.(\ref{wdf}) are valid 
in (\ref{wlc})
if the system operators pertain to a harmonic oscillator. When the system is 
nonlinear, as in the present study (and in the overwhelming majority of the 
cases in quantum and nonlinear optics), the usual practice is to assume that 
the dissipative terms remain unaffected by the Wigner higher derivative terms 
in Eq. (\ref{weq}). The validity of this assumption was examined earlier 
by Haake et. al. \cite{haake} and also by us \cite{gg2}. It is now well-known 
that the assumption is quite satisfactory in semiclassical limit and when 
nonlinearity is not too strong. Since the calculation of barrier crossing
dynamics needs the linearization of the potential at the barrier top as an
essential step, the nonlinearity of the potential $V(x)$ must be weak. This
assumption is essential for validity of Eq.(\ref{wlc}).
We also mention here that weakness of  nonlinearity is 
essential for the linearized description of the potential near the extrema
in almost all theories including path integral techniques upto date
\cite{woly,pollak1,miller2,hanggi}.

Second, the system-reservoir dynamics as governed by Eq.(\ref{wlc})  or 
(\ref{wdf}) is based on Markov approximation which implies that the underlying 
stochastic process due to reservoir is Markovian by construction as in 
classical theory. However in deriving Eq.(\ref{wlc}) or (\ref{wdf}) the 
system-reservoir coupling is assumed to be arbitrary \cite{leggett}. Thus 
Eq.(\ref{wlc}) is applicable to both in the strong and weak damping limit. 
Since we are considering here the spatial-diffusion limited rate we restrict 
ourselves from intermediate to strong damping range.

Eq.(\ref{wlc}) has been used earlier in several occasions. For example, Zurek 
and Paz \cite{zurek} have studied some interesting aspects of 
quantum-classical correspondence in relation to decoherence. Based on this 
equation a theory of quantum noise in classically chaotic dissipative systems 
has been formulated \cite{bag1,bag2}. The primary reason for choosing 
Eq.(\ref{wlc}) as our starting point is that it reaches the correct classical 
limit, i.e., the classical Kramers' equation when $\hbar \rightarrow 0$ and
$D$ becomes a thermal diffusion coefficient in the high temperature limit
and $W(x,v,t)$ reduces to classical probability distribution function in
phase space. Thus the WLC equation (\ref{wlc}) may be interpreted as a
semiclassical Kramers' equation.

Our object here is to search for a semiclassical solution in powers of $\hbar$
for the WLC equation (\ref{wlc}) under appropriate
boundary conditions and calculation of the barrier crossing rate in the 
intermediate to strong damping regime.


\section{The model, the stationary flux and the well population}

We now consider a particle of unit mass moving in a cubic potential $V(x)$
of the form
\begin{equation}
\label{potnl}
V(x) = -\frac{1}{3} A x^3 + B x^2 \; \; ,
\end{equation}

\noindent
where $x$ now corresponds to the reaction co-ordinate and its values at the 
extrema of the potential at $x=0$ and $x=x_0$ denote the reactant and the 
transition state, respectively. In this model all the remaining relevant 
degrees of freedom of the system and the environment constitute a heat bath
at a finite temperature $T$. Eq.(\ref{wlc}) with Eq.(\ref{potnl}) 
provides a complete description
of the stochastic process in terms of Wigner probability density function of
the $c$-number variables $x$, $v$.

One of the decisive advantage of using Wigner's phase space language is that 
it offers an excellent opportunity to extend the classical formulae to the
quantum domain when one replaces the classical probability functions by the 
Wigner density functions. Thus the original reasoning of Farkas and Kramers 
\cite{kramers,rmp1}
can be followed in the present problem to determine the steady state escape
rate from the well by considering a steady state situation in which a 
steady state probability current is maintained between the source and the
sink, the latter being situated beyond the transition state. The standard 
expression for escape rate in terms of flux over the population is thus 
given by
\begin{equation}
\label{rate}
k=\frac{j}{n_a}
\end{equation}

\noindent
where
\begin{equation}
\label{current}
j = \int_{-\infty}^{+\infty} dv \; v \; W(x_0, v)
\end{equation}

\noindent
and
\begin{equation}
\label{number}
n_a = \int_{left \; well} \int dx \; dv \; W(x,v)
\end{equation}

\noindent
and $W(x,v)$ is the steady state Wigner probability density function.


\section{The steady state semiclassical description : Boundary conditions}

The search for a semiclassical probability distribution function $W(x,v)$
for Eq.(\ref{wlc})
essentially rests on developing $W(x,v)$ in a power series of $\hbar$. Thus 
we write \cite{wigner}
\begin{equation}
\label{power}
W(x,v) = e^{-\beta \epsilon} \left [ \rho_0 (x,v) + \hbar \; \rho_1 (x,v) +
\hbar^2 \; \rho_2 (x,v) + \ldots \right ]
\end{equation}

\noindent
where $\beta = 1/k_BT$ and $\epsilon = \frac{1}{2} v^2 + V(x)$. As pointed
out by Wigner \cite{wigner}, $\rho_k$ will not involve higher derivatives 
of $V$ than $k$-th, nor higher powers of $v$ than $k$-th. This is relevant 
for the explicit form of $V(x)$ as given by Eq.(\ref{potnl}) in the present 
case and for calculation of higher terms in Eq.(\ref{power}) in a simpler
way.

By inserting (\ref{power}) in Eq.(\ref{wlc}) and equating the coefficients of 
different powers of $\hbar$ in the usual way and putting 
$\frac{\partial W}{\partial t} = 0 $ we obtain the following equations,
\begin{mathletters}
\begin{equation}
\label{rzero}
D \frac{\partial^2 \rho_0}{\partial v^2} + \left \{
\frac{\partial V}{\partial x} - 2\gamma v \right \} 
\frac{\partial \rho_0}{\partial v} - v
\frac{\partial \rho_0}{\partial x} = 0
\end{equation}

\begin{equation}
\label{rone}
D \frac{\partial^2 \rho_1}{\partial v^2} + \left \{
\frac{\partial V}{\partial x} - 2\gamma v \right \} 
\frac{\partial \rho_1}{\partial v} - v
\frac{\partial \rho_1}{\partial x} = 0
\end{equation}

\begin{eqnarray}
\label{rtwo}
D \frac{\partial^2 \rho_2}{\partial v^2} + \left \{
\frac{\partial V}{\partial x} - 2\gamma v \right \} 
\frac{\partial \rho_2}{\partial v} - v
\frac{\partial \rho_2}{\partial x} & = &
\frac{1}{4!} \frac{\partial^3 V}{\partial x^3} \left [ \; \left \{ \;
\frac{12 \gamma^2}{D^2} v - \frac{8 \gamma^3}{D^3} v^3 \; \right \} \rho_0
\right. \nonumber \\
& + & \left. \left \{ \; \frac{12 \gamma^2}{D^2} v^2 -\frac{6\gamma}{D} \;
\right \} \frac{\partial \rho_0}{\partial v} - \frac{6\gamma}{D} v 
\frac{\partial^2 \rho_0}{\partial v^2} + 
\frac{\partial^3 \rho_0}{\partial v^3} \right ]
\end{eqnarray}

\end{mathletters}

\noindent
and so on.

We note that the Eqs.(\ref{rzero}),(\ref{rone}) are homogenous partial 
differential equations while (\ref{rtwo}) is an inhomogenous one. The 
inhomogenous contribution begins to appear from $\rho_2$-term which is 
associated as a coefficient of $\hbar^2$ in the expansion (\ref{power}). The 
leading order quantum contribution (since $\rho_1$ does not contribute 
\cite{wigner}) through $\rho_2$ involves only cubic derivative of the 
potential in (\ref{rtwo}). The other derivatives of any arbitrary potential 
$V(x)$ contribute to the terms containing higher powers of $\hbar$ (e.g., 
$\hbar^4$, $\hbar^6$ etc.). Thus the leading order nonlinearity that affects 
the dynamics at the semiclassical level must at best be cubic one.

Eq.(\ref{rzero}) is the evolution of familiar {\it dynamical} contribution 
from classical Kramers' equation. The leading order quantum correction starts 
at $\hbar^2$ \cite{wigner}. Thus $W$ has the form
\begin{equation}
\label{wkd}
W(x,v) = e^{-\beta \epsilon} \left [ \; \rho_0 (x,v) + \hbar^2 \rho_2 (x,v)
+ \ldots \; \right ]
\end{equation}

\noindent
where $\rho_0$ and $\rho_2$ are the solutions of the equations (\ref{rzero})
and (\ref{rtwo}) respectively.

To obtain the steady state probability density $W(x,v)$ several requirements 
\cite{kramers} must be fulfilled:

(i) Near the bottom of the well all the particles are thermalized and are in a 
state of thermal equilibrium. In the classical  case the probability density 
is given by the Boltzmann distribution. In the present semiclassical case the 
corresponding distribution function contains an appropriate Wigner quantum
corrections of $\hbar^2$ order to Boltzmann distribution. According to Wigner the
semiclassical equilibrium distribution is given by
\begin{equation}
\label{wbd}
W_{BW}(x,v) = e^{-\beta \epsilon} + \hbar f_1 + \hbar^2 f_2 + \ldots
\end{equation}

\noindent
which is to be obtained as a solution of Eq.(\ref{weq}).

Following Wigner \cite{wigner} one can easily show that
\begin{equation}
\label{fone}
f_1 = 0
\end{equation}

\noindent
and
\begin{equation}
\label{ftwo}
f_2 = 4\pi^2 \; e^{-\beta \epsilon} \left [ \; -\frac{\beta^2}{8} 
\frac{\partial^2 V}{\partial x^2} + \frac{\beta^3}{28} \left (
\frac{\partial V}{\partial x} \right )^2 + \frac{\beta^3 v^2}{24}
\frac{\partial^2 V}{\partial x^2} \; \right ]
\end{equation}

We emphasize the distinguishing feature of the two expansions (\ref{wkd})
and (\ref{wbd}). While (\ref{wbd}) refers to an {\it equilibrium} thermal
Boltzmann-Wigner distribution, (\ref{wkd}) denotes a {\it steady state}
Wigner-Kramers distribution which takes into account of semiclassical
Brownian dynamics. (\ref{power}) may be viewed as a generalization of
Kramers' ansatz in the quantum mechanical context.

We make use of the solutions (\ref{wbd}-\ref{fone}) near the bottom of 
the well. We thus impose the following condition on $W(x,v)$ of (\ref{wkd})
as
\begin{mathletters}
\begin{equation}
\label{bc1}
W(x,v) \longrightarrow W_{BW} (x,v) \; \; \; \; {\rm as} \; \; x \; 
\approx \; 0 \; \; \; {\rm and \; all} \; v \; \; .
\end{equation}

This, in turn, implies $\rho_0(x,v) \rightarrow 1$ and
$\rho_2(x,v) \rightarrow e^{+\beta \epsilon} f_2$.

(ii) On the other hand beyond the transition state, i.e., $x=x_0$ all the
particles are removed by the sink, or,
\begin{equation}
\label{bc2}
W(x,v) \approx 0 \; \; \; \; {\rm as} \; \; x \; 
> \; x_0 \; \; \; {\rm for \; all} \; v \; \; .
\end{equation}
\end{mathletters}

Having specified the boundary conditions (\ref{bc1},\ref{bc2})
for Wigner distribution function
$W(x,v)$ we now elucidate the dynamics near the barrier at $x=x_0$. Near this
point we assume the usual linearized potential (with frequency $\omega_b$),
\begin{equation}
\label{potw}
V(x) = E_0 - \frac{1}{2} \omega_b^2 (x-x_0)^2
\end{equation}

\noindent
where $E_0$ is the barrier height and
\begin{equation}
\label{freq}
\omega_b^2 = -V''(x_0) \; > \; 0 \; \; .
\end{equation}

\noindent
The probability distribution function $W(x,v)$ is determined by 
$\rho_0(x,v)$ and $\rho_2(x,v)$ which obey the following stationary 
equations around the barrier ;
\begin{equation}
\label{steq1}
D\frac{\partial^2 \rho_0}{\partial v^2} + \left \{ -\omega_b^2 (x-x_0) -
2\gamma v \right \} \frac{\partial \rho_0}{\partial v} -
v \frac{\partial \rho_0}{\partial x} = 0
\end{equation}

\noindent
and
\begin{eqnarray}
\label{steq2}
& & D\frac{\partial^2 \rho_2}{\partial v^2} + \left \{ -\omega_b^2 (x-x_0) -
2\gamma v \right \} \frac{\partial \rho_2}{\partial v} - 
v \frac{\partial \rho_2}{\partial x} \nonumber \\
& = & -\frac{1}{12} \; A \;
\left [ \left \{ \; \frac{12\gamma^2}{D^2} v - \frac{8\gamma^3}{D^3} v^3 \;
\right \} \rho_0
+ \left ( \; \frac{12\gamma^2}{D} v - 6 \gamma \; \right )
\frac{\partial \rho_0}{\partial v} - 6\gamma v 
\frac{\partial^2 \rho_0}{\partial v^2} + D
\frac{\partial^3 \rho_0}{\partial v^3} \; \right ]
\end{eqnarray}

The inhomogenous term in Eq.(\ref{steq2}) requires the solution $\rho_0(x,v)$,
which can be obtained by solving (\ref{steq1}) directly using Kramers' method
\cite{kramers,rmp1} and is given by
\begin{equation}
\label{rhou}
\rho_0(u) = C_0 \int_0^u \exp \left \{ -\frac{\lambda}{2D} z^2 \right \} \;
dz + C_0^\prime
\end{equation}

\noindent
where $C_0$ and $C_0^\prime$ are constants to be determined. Here
\begin{equation}
\label{uvax}
u = v + a (x-x_0)
\end{equation}

\noindent
where $a$ is a solution of $a^2 +2\gamma a - \omega_b^2 =0$ and 
$\lambda = -(2\gamma +a)$. The relevant root of $a$ which makes $\lambda > 0$
is $a_- = -\gamma - \sqrt{\gamma^2+\omega_b^2})$, so that 
$\lambda = -\gamma + \sqrt{\gamma^2+\omega_b^2}$. The requirement that
\begin{equation}
\rho_0 \longrightarrow 0 \; \; \; {\rm as} \; \;  x \; \rightarrow \; \infty
\end{equation}

\noindent
implies $C_0^\prime = C_0 \sqrt{\frac{\pi D}{2\lambda} }$. Thus
\begin{equation}
\label{ksoln}
\rho_0(u) = C_0 \left [ \; \sqrt{\frac{\pi D}{2\lambda} } + \int_0^u
\exp \left \{ - \frac{\lambda}{D} z^2 \right \} \; dz \; \right ]
\end{equation}

\noindent
Eq.(\ref{ksoln}) is essentially the Kramers' solution which we now use to 
determine the inhomogenous terms in equation for $\rho_2$ (Eq.(\ref{steq2})).
Three pertinent points are to noted here. Since we are working in the 
semiclassical limit $k_BT > \hbar \omega$, we keep the terms upto the
${\cal O} \left[ \left ( \frac{1}{k_BT} \right )^2 \right ]$ in the
inhomogenous part. Second, we take into account of the Wigner's remark 
\cite{wigner} on
quantum contribution contained in $\rho_2$ term. The term has been 
``interpreted as meaning that a quick variation of the probability function
with co-ordinates is unlikely, as it would mean a quick variation, a short
wavelength, in the wave functions. This however would have the consequence
of high kinetic energy''. The finite range of co-ordinates therefore ranges 
over $\sim \hbar/{\overline v}$ where ${\overline v}$ is mean momentum
$\sim (k_BT)^{1/2}$. Thirdly, we also note that the leading order
nonlinearity enters through the cubic derivative of the potential 
(\ref{potnl}) in the inhomogenous term.

Based on these consideration we are now led to the following equation for 
$\rho_2$ :
\begin{equation}
\label{mrho}
\frac{d^2 \rho_2}{d u^2} + \frac{\lambda}{D} u \frac{d \rho_2}{d u} =
( \; - M_1 - M_2 u - M_3 u^2 \; ) \; \exp\left (-\frac{\lambda u^2}{2D} 
\right ) \; \; ,
\end{equation}

\noindent
where the transformation of derivatives has been carried using the change of
variables through the linear combination of $x$ and $v$ as expressed in
(\ref{uvax}). $M_1$, $M_2$ and $M_3$ are given by
\begin{mathletters}
\begin{eqnarray}
M_1 & = & A \; C_0 \left \{ \; \frac{\gamma^2}{D^3} (k_BT) -
\frac{\gamma}{2D^2} -\frac{\lambda}{12D^2} \; \right \} \; \; ,\\
M_2 & = & A \; C_0 \; \frac{\gamma \lambda}{2D^3} (k_BT)^{1/2} \\
{\rm and} \; \; \; 
M_3 & = & A \; C_0 \; \frac{\lambda^2}{12D^3}
\end{eqnarray}
\end{mathletters}

\noindent
It may be noted that all the three terms in the source term of Eq.(\ref{mrho})
are within $(1/k_BT)^2$ order (This may be checked by noting that $u$ behaves 
as $v$ near $x=x_0$ and ${\overline v} \sim (k_BT)^{1/2}$ as argued 
earlier).

The general solution of Eq.(\ref{mrho}) subject to boundary condition,
\begin{equation}
\label{gbc}
\rho_2 \longrightarrow 0 \; \; \; \; {\rm as} \; \; \; x \rightarrow \infty
\end{equation}

\noindent
is given by
\begin{eqnarray}
\label{rtsol}
\rho_2(u) & = & \left [ \; \frac{M_1 D}{\lambda} + \frac{M_2 D}{2\lambda} \; u
+ \frac{M_3 D}{3\lambda} \; u^2+ \frac{2}{3} \frac{M_3 D^2}{\lambda^2} \; 
\right ]
\exp\left ( -\frac{\lambda u^2}{2D} \right ) \nonumber \\
& & + \left ( C_2 - \frac{M_2 D}{2 \lambda} \right ) \int_0^u
\exp\left ( -\frac{\lambda z^2}{2D} \right ) \; dz + 
\sqrt{ \frac{\pi D}{2 \lambda} } \left ( C_2 - \frac{M_2 D}{2 \lambda} \right)
\; \; ,
\end{eqnarray}

\noindent
where $C_2$ is an integration constant to be determined and
$u = v - |a_-|(x-x_0)$ with $|a_-| = \gamma + \sqrt{\gamma^2 +\omega_b^2}$.
Thus upto $\hbar^2$ our distribution function (\ref{wkd}) becomes
\begin{equation}
\label{wusol}
W(u) = e^{-\beta \epsilon} \; [ \; \rho_0(u) + \hbar^2 \rho_2(u) \; ] \; \; ,
\end{equation}

\noindent
where $\rho_0(u)$ and $\rho_2(u)$ are given by (\ref{ksoln}) and 
(\ref{rtsol}), respectively.

To determine the two unknown constants $C_0$ and $C_2$ we now employ the
boundary condition (\ref{bc1}), which implies that the probability density
function $W(u)$ of (\ref{wusol}) must coincide with the thermal
Boltzmann-Wigner distribution $W_{BW}(x,v)$ of Eq.(\ref{wbd}) near $x=0$.

Thus we have from (\ref{wusol}) [i.e., from (\ref{ksoln}) and (\ref{rtsol})]
\begin{equation}
\label{left1}
W(x,v) \stackrel{x \rightarrow -\infty}{\longrightarrow} e^{-\beta \epsilon}
\; \left [ \; C_0 \sqrt{ \frac{2\pi D}{\lambda} } + \hbar^2 \left (
C_2 - \frac{M_2D}{2\lambda} \right )\sqrt{ \frac{2\pi D}{\lambda} } \; 
\right ]
\end{equation}

\noindent
on the other hand $W_{BW}(x,v)$ of (\ref{wbd}) [ with (\ref{fone}) and
(\ref{ftwo})] near the bottom of the well at $x=0$ where the linearized 
potential
\begin{equation}
\label{linearv}
V(x) = \frac{1}{2} \omega_0^2 x^2
\end{equation}

\noindent
may be used, behaves as the Boltzmann-Wigner distribution of the particles 
at thermal equilibrium ;
\begin{equation}
\label{left2}
W_{BW}(x,v) \stackrel{{\rm Near} \; x=0}{\longrightarrow} 
e^{-\beta \epsilon} \; \left [ \; 1 - 
\frac{1}{3} \hbar^2 \pi^2 \beta^2 \omega_0^2 \; \right ] \; \; .
\end{equation}

\noindent
Matching of (\ref{left1}) and (\ref{left2}) suggests that
\begin{mathletters}
\begin{equation}
\label{c0val}
C_0 = \sqrt{ \frac{\lambda}{2\pi D} }
\end{equation}

\noindent
and
\begin{equation}
\label{c2val}
\left ( \frac{M_2 D}{2\lambda} - C_2 \right ) = \frac{1}{3}
\sqrt{ \frac{\lambda}{2\pi D} } \; \pi^2 \beta^2 \omega_0^2 \; \; .
\end{equation}
\end{mathletters}

\noindent
The boundary condition (\ref{bc1}) therefore uniquely determines the
probability distribution function (\ref{wusol}) where $\rho_0$ and $\rho_2$
are given by (\ref{ksoln}) and  (\ref{rtsol}) and $C_0$ and $C_2$ by
(\ref{c0val},\ref{c2val}) respectively. The resultant unnormalized 
expression is as follows:
\begin{eqnarray}
\label{solution}
W(x,v) & = & e^{-\beta \epsilon} \left [ \; \sqrt{ \frac{\lambda}{2\pi D} }
\; \left \{ \; \sqrt{ \frac{\pi D}{2 \lambda} } + \int_0^u
\exp\left ( -\frac{\lambda}{D} z^2 \right ) \; dz \; \right \}  \right.
\nonumber \\
& & + \hbar^2 \; \left \{ \; \exp\left ( -\frac{\lambda}{D} u^2 \right ) \;
\left ( \; \frac{M_1 D}{\lambda} + \frac{M_2 D}{2\lambda} \; u +
\frac{M_3 D}{3\lambda} \; u^2 + \frac{2}{3} \frac{M_3 D^2}{\lambda^2} \; 
\right ) \right. \nonumber \\
& & \left.  \left. 
- \frac{1}{3} \sqrt{ \frac{\lambda}{2\pi D} } \; \pi^2 \; \beta^2 
\; \omega_0^2 \int_0^u \exp\left ( -\frac{\lambda}{2D} z^2 \right ) \; dz \;
- \frac{1}{6} \pi^2 \; \beta^2 \; \omega_0^2 \; \right \} \; \right ]
\end{eqnarray}

\noindent
with $ u = v - |a_-|(x-x_0)$.


\section{ The escape rate}

In order to determine the semiclassical rate of escape we now make use of
the flux-over-population formula (\ref{rate}). To this end we obtain the
well population from Eqs.(\ref{number}) and (\ref{solution}), assuming the
linearized potential $V(x)=\frac{1}{2} \omega_0^2 x^2$ near the bottom of
the well at $x=0$.
\begin{equation}
\label{newnum}
n_a = \frac{2\pi}{\omega_0 \beta} \left [ \; 1 - \frac{1}{3} \hbar^2
\pi^2 \beta^2 \omega_0^2 \; \right ] \; \; .
\end{equation}

\noindent
The probability current (\ref{current}) becomes
\begin{eqnarray}
\label{newcurrent}
j & = &  e^{-\beta E_0} \left [ \; \sqrt{ \frac{\lambda}{2\pi D} } \; 
\frac{1}{\beta} \; 
\sqrt{ \frac{\pi}{ \frac{\beta}{2} + \frac{\lambda}{2D} } } \right. \nonumber\\
& & \left. + \; \hbar^2 \left \{
\; \frac{M_2 D}{2 \lambda} \; \frac{1}{ \frac{\lambda}{D} +\beta } \;
\sqrt{ \frac{\pi}{ \frac{\beta}{2} + \frac{\lambda}{2D} } } 
- \frac{1}{3} \; \sqrt{ \frac{\lambda}{2\pi D} } \; 
\pi^2 \; \omega_0^2 \; \beta^2 \; \frac{1}{\beta} \;
\sqrt{ \frac{\pi}{ \frac{\beta}{2} + \frac{\lambda}{2D} } } \; \right \} \;
\right ] \; \; ,
\end{eqnarray}

\noindent
where we have assumed the linearized potential 
$V(x) = E_0 - \frac{1}{2} \omega_b^2 (x-x_0)^2$ at the barrier top in the 
expression for probability density function (\ref{solution}).

It is thus interesting to note that the quantum correction to classical
population and flux at the semiclassical level ($\hbar^2$-order). These yield
the expression for rate of escape (\ref{rate}) as
\begin{equation}
\label{newrate}
k = e^{ - E_0/k_B T } \; \frac{\omega_0}{2\pi} \; 
\frac{ \sqrt{\gamma^2+\omega_b^2}-\gamma}{\omega_b}  \; 
\left [ \; 1 + \hbar^2 \; \frac{A}{8 \; (k_BT)^{3/2} } \; 
\frac{1}{ \sqrt{\gamma^2+\omega_b^2}+\gamma} \; 
\right ] \; \; ,
\end{equation}

\noindent
where we have put $\beta = 1/k_BT$, 
$\lambda = -\gamma + \sqrt{ \gamma^2 +\omega_b^2}$ and $A$ is the nonlinear
parameter of the cubic potential (\ref{potnl}).

The above expression can be rewritten as
\begin{equation}
k = k_{cl} \; + \; k_{semi} \; \; ,
\end{equation}

\noindent
where $k_{cl}$ and $k_{semi}$ correspond to classical and semiclassical Kramers'
rate, respectively, with
\begin{mathletters}
\begin{eqnarray}
k_{cl} & = & \frac{\omega_0}{2\pi} \; 
\frac{ \sqrt{\gamma^2+\omega_b^2}-\gamma}{\omega_b}  \; 
e^{- E_0/k_BT} \; \; ,\\
k_{semi} & = & \frac{\omega_0}{2\pi \omega_b} \; 
\left ( \frac{A \hbar^2}{8 \; (k_BT)^{3/2} } \right ) \;
\left ( \frac{ \sqrt{\gamma^2+\omega_b^2}-\gamma }
{ \sqrt{\gamma^2+\omega_b^2}+\gamma} \right ) \;  
e^{- E_0/k_BT} \; \; .
\end{eqnarray}
\end{mathletters}

\noindent
The result (\ref{newrate}) describes the spatial diffusion controlled rate of
escape at moderate to strong friction $\gamma$ at the semiclassical level
where the leading order quantum correction makes its presence felt. 
We also emphasize that since we are using high temperature 
($ k_BT > \hbar \omega$) semiclassical 
surrounding due to Caldeira and Leggett, the result cannot be extrapolated 
to zero temperature/low temperature limit, when non-Markovian effects also 
begin to influence the dynamics. The temperature dependence of 
$\hbar^2$-containing term in (\ref{newrate}) signifies that as $k_BT$ 
becomes large the quantum contribution significantly diminishes - a typical
semiclassical feature in the dynamics, as expected. For
$\gamma=0$ the expression (\ref{newrate}) yields the result of ``simplest''
transition state theory $k_{TST}$ which includes the quantum correction
at the semiclassical as well
\begin{equation}
\label{ktst}
k_{TST} = \frac{\omega_0}{2\pi} \; 
\left \{ \; 1 \; + \; \frac{\hbar^2 A}{8 \; \omega_b \; (k_BT)^{3/2} }  
\; \right \} \; e^{ - E_0/k_B T } \; \; .
\end{equation}

For the strong friction case, Eq.(\ref{newrate}) can be simplified further 
to give the rate of escape in the overdamped regime ($\gamma \gg \omega_b$),
\begin{equation}
\label{kodamp}
k_{overdamped} = \frac{\omega_0 \omega_b}{2\pi} \; 
\left \{ \; \frac{1}{\gamma} \;  + \; 
\frac{\hbar^2 A}{32 \; \gamma^2 \; (k_BT)^{3/2} }  
\; \right \} \; e^{ - E_0/k_B T } \; \; ,
\end{equation}

\noindent
which approaches to zero as $\gamma \rightarrow \infty$. $\hbar^2$ term in the
above expression reveals a stronger friction ($1/\gamma^2$) dependence, which
implies that quantum correction term is more susceptible to damping. The
classical limit of (\ref{kodamp}) is the Smoluchowski limit. Therefore the 
result (\ref{kodamp}) may be regarded as the semiclassical
Smoluchowski-Kramers limit.


\section{Conclusions}

The main purpose of this paper is to explore whether a differential equation
(for probability distribution function) based approach is possible for
calculation of quantum Kramers' rate. We show that 
Wigner-Leggett-Caldeira [WLC] equation which 
describes the quantum Brownian motion in a force field, in a high temperature
Ohmic environment is suitable for this purpose. Based on an expansion in 
powers of $\hbar$ we solve this equation
for Wigner probability density function of $c$-number variables under
appropriate boundary conditions. We make use of Wigner's quantum correction 
to Boltzmann factor to describe the thermalization of the particles deep
inside the well. The effective dynamics is spatial-diffusion 
limited and the drift term ($\gamma$) is Markovian. The validity of the
theory is ensured at high temperature and for the intermediate to strong 
damping regime, where the
quantum corrections upto the order $\hbar^2$ significantly modifies the
classical Kramers' results.

We now summarize the main conclusions of the study:

(i) We have identified WLC equation as a semiclassical Kramers' equation and
calculated explicitly the barrier crossing dynamics. The method is closer to
the classical one and is different from the existing path integral approaches
for calculation of quantum Kramers' rate. While the existing procedure to
calculate the total rate is to add $k_{quant}$ to the classical Kramers' rate
$k_{cl}$, the present method yields the total semiclassical rate. In the limit
$\hbar/k_BT \rightarrow 0$ the classical rate is recovered.

(ii) The quantum correction to classical Kramers' rate includes 
$1/\gamma^2$ dependence in the overdamped limit ($\gamma \rightarrow 
\infty$).

(iii) The semiclassical theory can be reduced to `simplest' quantum
transition state description for $\gamma=0$ situation. The rate
becomes susceptible to the ratio of nonlinear coefficient to linear barrier
frequency.

(iv) The role of nonlinearity gets entangled with quantum correction
in this description. This is obvious from the appearance of the higher
derivatives of the potential in the Wigner equation. The quantum noise 
term appears as a source term in classical equation.

(v) Kramers' celebrated ansatz in classical theory (the factorization of an 
equilibrium factor from a dynamical factor) is generalized in the present 
semiclassical context.

(vi) Compared to other existing methods the quantum-classical correspondence
in the problem of barrier crossing dynamics is relatively more transparent
in the present approach.

In view of the considerable development of the theory in the weak
coupling limit, particularly in the quantum optical context,
we believe that the Wigner function approach like the present one can
be suitably extended to analyze the energy diffusion rate in the 
semiclassical domain in the Markovian limit. Since at low temperatures,
the non-Markovian effects begin to be appreciable, one feels the need
for appropriate non-Markovian quantum Brownian equations. In absence of these 
descriptions
any differential equation based approach is to remain far from promising.

\acknowledgments
Partial financial support from Department of Science and Technology, 
Government of India is thankfully acknowledged. BCB is thankful to the 
Council of Scientific and Industrial Research, Government of India for a 
fellowship.


\end{document}